\documentclass[preprint,twocolumn,12pt,5p]{elsarticle} 
\usepackage[utf8]{inputenc}
\usepackage{graphicx}
    \graphicspath{ {Figures/} }
\usepackage{float}
\usepackage{caption}                            
\usepackage{amsmath, amsthm, amssymb}
\usepackage{textgreek}
\usepackage{mathtools}
\usepackage{gensymb}
\usepackage[version=4]{mhchem}
\usepackage[linktoc=all,pdfencoding=auto, psdextra]{hyperref}
\hypersetup{pdfauthor=author}
\usepackage{lineno}
\usepackage{orcidlink}
\usepackage{natbib}

\usepackage[normalem]{ulem} 

\newcommand{\bNMR}{\textbeta-NMR\ }
\newcommand{\btext}{\textbeta\ }

\newcommand{\sigmap}{\textsigma\textsuperscript{+}}

\journal{Physica Scripta}

\date{\today}

\begin{document}

    \begin{frontmatter}
    \title{Fully upgraded \textbeta-NMR setup at ISOLDE for high-precision high-field studies}
\author[CERN,Darmstadt]{M. Jankowski\orcidlink{0009-0004-6800-6190}\corref{cor1}}
    \ead{marcus.jankowski@cern.ch}
    \cortext[cor1]{Corresponding authors}
\author[CERN,Poznan]{N. Azaryan}
\author[CERN,Poznan]{M. Baranowski\orcidlink{0000-0002-8584-7029}}
\author[CERN,Manchester]{M. L. Bissell\orcidlink{0000-0002-0144-5015}}
\author[GSI]{H. Brand}
\author[CERN,UniGe]{M. Chojnacki\orcidlink{0000-0002-9733-0529}}
\author[CERN,UniGe]{J. Croese\orcidlink{0000-0002-0660-4409}\fnref{fn0}}
    \fntext[fn0]{Current address: Netherlands Organisation for Applied Scientific Research (TNO), Delft, Netherlands}
\author[CERN,Leipzig]{K. M. Dziubinska-K{\"u}hn\orcidlink{0000-0002-4526-841X}\fnref{fn1}}
    \fntext[fn1]{Current address: Maastricht University, Maastricht, Netherlands}
\author[UniGe]{B. Karg\orcidlink{0000-0003-2648-3509}}
\author[Tennessee]{M. Madurga Flores\orcidlink{0000-0002-8177-4328}}
\author[Helsinki]{M. Myllym{\"a}ki}
\author[CERN]{M. Piersa-Silkowska\orcidlink{0000-0002-5877-2818}}
\author[CERN]{L. Vazquez Rodriguez\orcidlink{0000-0002-0093-2110}\fnref{fn2}}
    \fntext[fn2]{Current address: MPIK, Heidelberg, Germany}
\author[UniGe]{S. Warren\orcidlink{0000-0003-2429-6726}}
\author[Czech]{D. Zakoucky}
\author[CERN,UniGe]{M. Kowalska\orcidlink{0000-0002-2170-1717}\corref{cor1}}
    \ead{magdalena.kowalska@cern.ch}
\address[CERN]{CERN, Geneva, Switzerland}
\address[Darmstadt]{TU Darmstadt, Darmstadt, Germany}
\address[Poznan]{Adam Mickiewicz University, Poznan, Poland}
\address[Manchester]{University of Manchester, Manchester, United Kingdom}
\address[GSI]{GSI, Darmstadt, Germany}
\address[UniGe]{University of Geneva, Geneva, Switzerland}
\address[Leipzig]{Leipzig University, Leipzig, Germany}
\address[Helsinki]{University of Helsinki, Helsinki, Finland}
\address[Tennessee]{University of Tennessee, Knoxville, USA}
\address[Czech]{Czech Academy of Sciences, Rez, Czech Republic}

\begin{abstract}

\bNMR is an advancing technique that enables measurements relevant to various fields of research, ranging from physics to chemistry and biology.
Among the recent achievements of the \bNMR setup located at the ISOLDE facility at CERN is the determination of the magnetic moment of a short-lived nucleus with a part-per-million accuracy.
Presented here are major upgrades and extensions of that \bNMR setup.
The most important advancement is the installation of a 4.7~T superconducting solenoidal magnet, leading to sub-ppm spatial homogeneity and temporal stability of the magnetic field.
A detector array optimised for such magnetic field has also been implemented and a more powerful, time-resolved, fully-digital data acquisition system has been deployed. 
To commission the upgraded beamline, \bNMR resonances of short-lived \textsuperscript{26}Na were recorded in solid and liquid samples. 
These showed 3-fold narrower linewidths and 15-fold higher resolving power than using the previous setup.
Hence, the improvements achieved here permit more accurate bio-\bNMR studies, investigating, e.g.,\ the interaction of metal ions with biomolecules, such as DNA.
They also pave the way for the first studies of the distribution of the magnetisation inside short-lived nuclei.

\end{abstract}

    \end{frontmatter}
    

\section{Introduction}

Nuclear magnetic resonance (NMR) is a powerful spectroscopic technique used in physics, chemistry and biology \cite{Mazzola03}.
It relies on the interaction of nuclear spins with strong static and weak oscillating magnetic fields.
NMR can yield information on the probed nuclei or their environment \cite{Mazzola03, Abragam1989} through the nuclei's Larmor frequency and relaxation time that reflects how fast they return to thermal equilibrium.

Unfortunately, conventional room-temperature NMR suffers from very low sensitivity. 
This is caused by low polarisation, i.e.\ a small population difference in spin orientations, as well as inefficient signal detection via electromagnetic induction \cite{Levitt:500323}.
One way to enhance the sensitivity is to increase the polarisation above room temperature equilibrium, known as hyperpolarisation.
Different techniques are used to achieve hyperpolarisation in stable nuclei, e.g.\ dynamic nuclear polarisation, parahydrogen-induced polarisation, or optical pumping \cite{Ellis2023}. 

The use of short-lived \textbeta-emitting probe nuclei represents another sensitivity gain for NMR, since it is based on the detection of the emitted \btext particles.
Specifically, due to the parity non-conservation of the weak interaction, spin-polarised radioactive isotopes exhibit an anisotropy in their \btext decay \cite{Wu1957}.
This anisotropy is destroyed if radio-frequency (rf) excitations are applied around the Larmor frequency of the nucleus. By recording the asymmetry as a function of the excitation frequency, an NMR spectrum can be recorded on as few as a million nuclei.
The main application of \textbeta-detected NMR (\textbeta-NMR) has so far been in nuclear physics, where it has allowed to determine magnetic dipole moments of unstable nuclei with per-cent or per-mill precision in solid hosts \cite{Kowalska2008} and electric quadrupole moments \cite{Keim2000, Yordanov2019}.

Furthermore, moving \bNMR towards liquid samples is a step forward in the achievable precision, as the molecular tumbling in the liquid diminishes anisotropic effects typically occurring in solid samples, resulting in narrower resonances \cite{Sugihara2017, Mihara2019, Croese2021}.
The combination of \textbeta-NMR's ultra-sensitivity with liquid samples can thus enable novel applications in more research fields.
First, magnetic moments of unstable nuclei determined with parts-per-million (ppm) accuracy and precision combined with precise hyperfine structure constants can yield information on the distribution of the neutrons in the nucleus through the so-called hyperfine anomaly \cite{BohrWeisskopf1950, Bissell2023}.
The second potential application lies in biochemistry studies that require measuring small shifts in the Larmor frequency (chemical shifts) of metal ions depending on their interaction with biomolecules in the sample, such as proteins or DNA \cite{Jancso2017}. 

A recent achievement of the technique has been the use of liquid hosts to determine the magnetic moment of a short-lived nucleus with 7~ppm accuracy \cite{Harding2020MagneticBiology} at this \bNMR setup at the ISOLDE facility at CERN. 
However, the accuracy and resolution need to be further improved to facilitate the planned future experiments: 1) the determination of the hyperfine anomaly in light nuclei, such as the well-known one-neutron halo nucleus \textsuperscript{11}Be \cite{Bissell2023} and 2) studies of the interaction of lighter alkali metals, Na and K, with DNA G-quadruplex structures \cite{Karg2020}. 

This contribution gives an overview on the major upgrades that have allowed to increase the precision and resolution achievable with the \bNMR beamline at ISOLDE to enable the above-mentioned studies.
These improvements include a 4.7~T superconducting magnet with sub-ppm homogeneity, a new magnetic field-compatible \textbeta-detector array, and a versatile and time-resolved digital data acquisition system.
The results of the commissioning experiment using \textsuperscript{26}Na in a liquid is compared to the spectra recorded before the upgrades \cite{Harding2020MagneticBiology, Croese2021}.

\section{Upgraded \bNMR beamline}
\label{section:Setup}

\begin{figure*}
    \centering
\includegraphics[width=\textwidth]{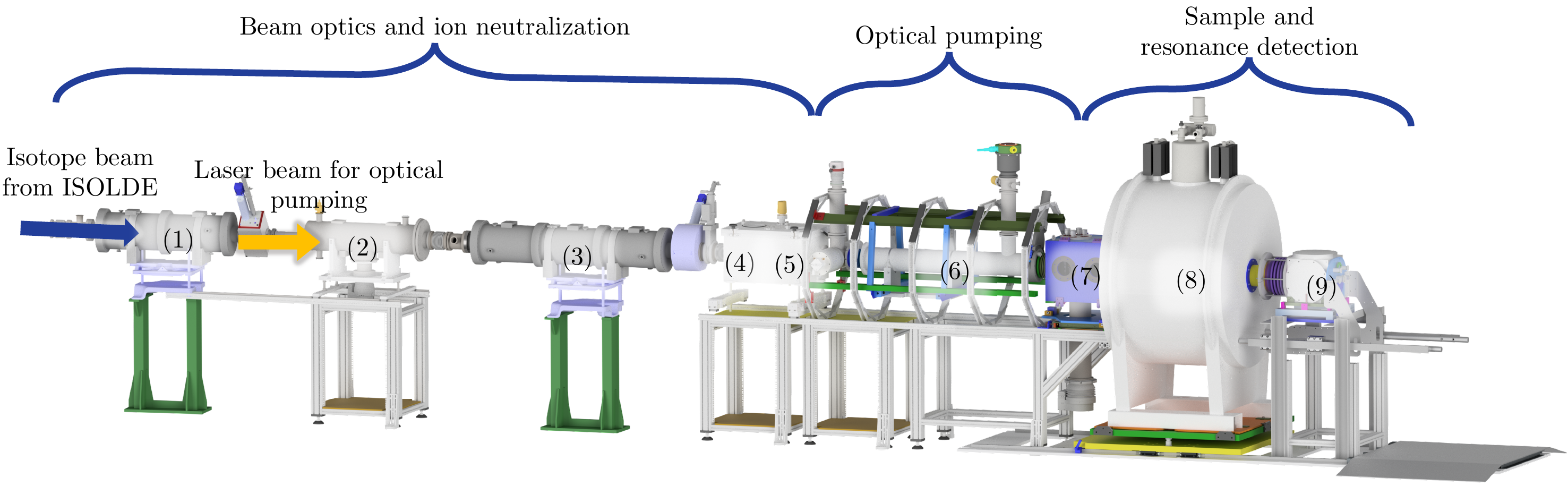}
    \caption{Overview on the components of the upgraded \bNMR beamline: (1) quadrupole doublet, (2) bender, (3) quadrupole triplet, (4) Doppler tuning electrode, (5) charge exchange cell, (6) optical pumping section, (7) front detector, (8) superconducting magnet, (9) retractable sample carriage.}
\label{fig:beamline}
\end{figure*}

The setup presented in this manuscript is designed to perform \bNMR experiments on optically-pumped beams of short-lived nuclei, as described in \cite{Croese2021, Gins2019} and \cite{Kowalska2017}.
A layout of the beamline and its upgraded components is depicted in Figure \ref{fig:beamline}.

The ion beam coming from ISOLDE enters the beamline through an electrostatic quadrupole doublet focusing element (1), followed by a bender (2). 
Here, the ion beam is deflected by 5\degree\, and thus brought to collinearly overlap with the laser beam used for optical pumping. 
The ion beam consequently passes through a quadrupole triplet (3) for further beam tuning.
Next, it reaches the Doppler-tuning electrodes (4) that bring the beam into resonance with the laser in the following optical-pumping section.
The beam then passes through the charge exchange cell (5) where it is optionally neutralised when propagating through Na vapour \cite{Croese2021, Gins2019}.
The optical pumping takes place in the subsequent 2-m long section, (6) before the polarised beam passes through the front detector (7) and is implanted into the sample located in the centre of the superconducting magnet (8). 
A rail system allowing to easily open the setup for a sample change is located at the end of the beamline (9).


\subsection{Superconducting magnet}
\label{section:SCmagnetBreadboard}

The most substantial change to the \bNMR beamline at ISOLDE is the installation of a new magnet. 
Previously, the setup used a 1.2~T electromagnet (Bruker BE10) with shimming coils.
This allowed for a magnetic field homogeneity of 4~ppm across the liquid sample support \cite{Croese2021}, which is a thin, 8~mm diameter mica disk placed at 45\degree\ with respect to vertical and to the beam direction.
In comparison, the new superconducting magnet (Bruker 47/16), see (8) in Figure \ref{fig:beamline}, has a magnetic field strength of 4.7~T. 
It has passive shimming and its magnetic-field homogeneity is in the sub-ppm range in the centre of the magnet, with deviations of less than 0.8~ppm in a sphere of 30~mm radius around the sample.
Additionally, the magnet is characterised by a high temporal field stability with a drift below 0.06~ppm/h.
The above features allow to achieve narrower NMR resonances, to increase the reproducibility of the measured Larmor frequencies, and to increase the spectral resolution.

\begin{figure} 
   \centering
   \includegraphics[width=0.9\columnwidth]{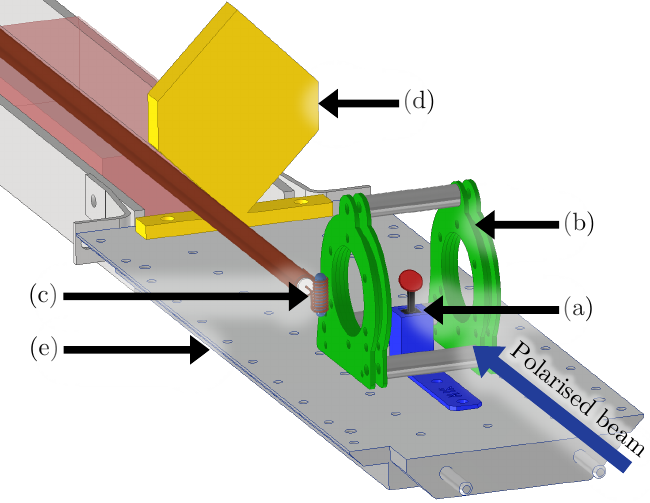}
   \caption{Schematic of the sample carriage and its components: (a) sample holder with sample, (b) rf coil, (c) reference probe, (d) rear \btext detector, (e) sample carriage. The polarised beam travels along the magnetic field axis from the bottom right and it gets implanted into sample.}
   \label{fig:SampleCarriage}
\end{figure}

The magnet has a 160~mm-wide bore that accommodates the last beamline section with the sample in its centre. 
Further components located inside the magnet, shown in Figure~\ref{fig:SampleCarriage}, include (a) the sample holder, (b) the rf coil for exciting spins of the probe nuclei, (c) the reference probe, and (d) the rear \btext detector behind the sample, which is described in Section \ref{section:Detectors}.
All of these components are fixed on an aluminium sample carriage (e), which rests on a rail system, allowing for an easy retraction from the centre of the magnet.

The reference probe (c) is used to perform pulsed NMR on \textsuperscript{2}H nuclei to determine the magnetic field $B_0$ at the time of the \bNMR measurements.
It consists of a 20-mm long sealed glass tube with 2~mm internal and 3~mm external diameter and an rf coil for transmission and pick-up of the signal.
The probe is located at the same height as the \bNMR sample, 25~mm from the beam axis, adjacent to the main rf coil.

When the sample carriage is inserted into the magnet, conical pins align it to the rest of the setup, thus ensuring a fast and highly reproducible positioning of all elements on the carriage with reference to the beam axis and each other.
This diminishes the uncertainty in the difference of the magnetic field between sample and reference NMR probe positions.

\begin{figure}
   \centering
   \includegraphics[width=\columnwidth]{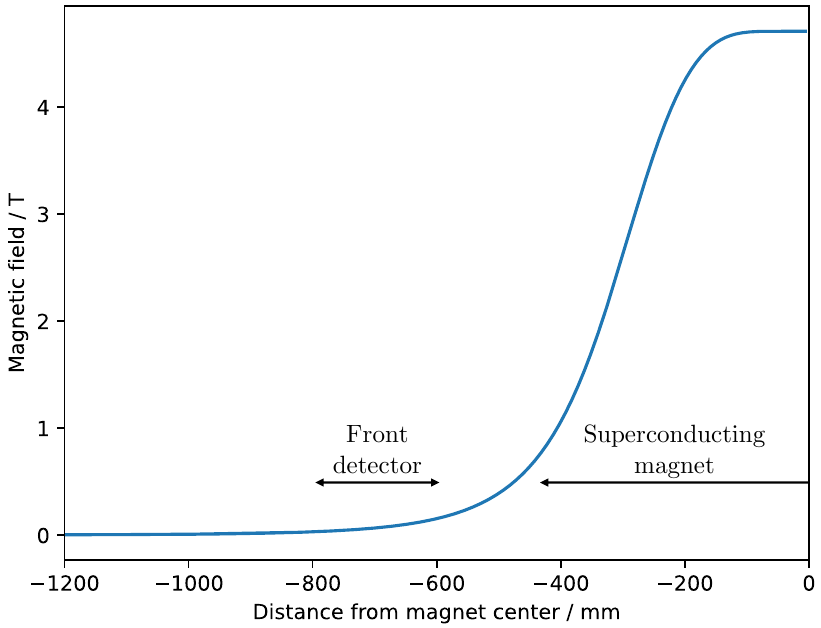}
   \caption{Magnetic field on the beam axis in the region between the end of the optical pumping section and the magnet's center. The front detector is located at -780~mm to -580~mm from the centre of the magnet.}
   \label{fig:magnetic-field}
\end{figure}

Another important consideration for \bNMR on optically-pumped beams is the magnetic field strength and direction, and its change in the transitional region between the optical pumping area and the magnet's centre.
In the former, four Helmholtz coils provide a 2-mT guiding field pointing in the direction of the magnet.
To ensure an efficient transfer of atomic spin polarisation to the nuclei during the decoupling of the electron and nuclear spin, the field strength and direction, and their change are important.
Previously, the magnetic field of the electromagnet was perpendicular to the beam and polarisation axis, which necessitated a slow rotation of the spins that led to inevitable losses in spin polarisation \cite{Croese2021}.
The field of the new superconducting magnet is along the polarisation axis and in the same direction as the guiding field, which avoids polarisation losses because no spin rotation is required.
The resulting overall magnetic field is shown in Figure \ref{fig:magnetic-field}.

\subsection{\btext detectors}

\label{section:Detectors}

When the implanted unstable nuclei decay in the sample, the 4.7~T magnetic field diverts the path of the emitted MeV \btext particles.
They propagate on spiral trajectories, whose gyro-radii are in the order of several~mm, depending on their angle of emission and energy.
As a result, \btext particles emitted at 0\degree\ to 90\degree\ with respect to the magnetic field and beam axes are guided in the forward direction, and these emitted at angles between 90\degree\ and 180\degree\ move backwards.

\begin{figure}
    \centering
    \includegraphics[width=0.9\columnwidth]{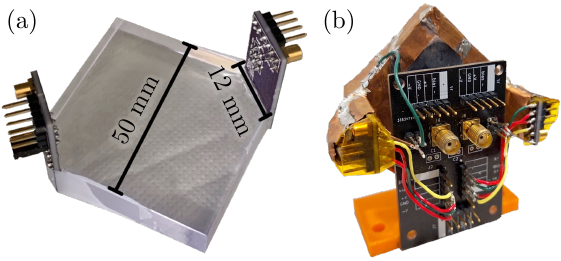}
    \caption{Rear detector. (a) Scintillator and PCBs with SiPMs and amplifiers on opposite sides. (b) Wrapped in aluminised Mylar, with breakout board for simple cable management in the middle and the two PCBs either side.}
    \label{fig:rear-detector}
\end{figure}

Therefore, the two \btext detectors were designed to be implemented along the magnetic field axis, one positioned in front of the sample and the other behind it.
They are made of EJ-200 plastic scintillators, whose optical signal is read out by 6 x 6 mm Onsemi 60035 silicon photomultipliers (SiPMs). 
The SiPMs are bonded to a printed circuit board with Texas Instruments OPA656 operational amplifiers for adjustable gain of the signals while minimising noise.
This results in signals with a full width half maximum (FWHM) of about 100~ns leading to possible rates below $10^6$~events/s per SiPM.
In addition, the detectors need to be light-tight to prevent background from the laser beam used for optical pumping.
Therefore, the scintillators are wrapped in a 50-\textmu m thick layer of aluminised Mylar.
As SiPMs are not affected by a magnetic field, the detector assemblies can be located in the region of the sample in the superconducting magnet without the need for light guides. 
The rear detector has dimensions of 50~mm by 50~mm and 10~mm thickness.
Two SiPMs are coupled to the scintillator on the opposite truncated corners, as shown in Figure \ref{fig:rear-detector}. 
The detector is located 170~mm behind the sample in the magnet's center, see Figure \ref{fig:SampleCarriage}. 

The front detector requires an aperture to let the polarised beam pass towards the sample.
Its design was guided by simulations using COMSOL and GEANT4 to optimise its angular acceptance for \btext particles emitted close to the beam axis, because these particles exhibit the highest emission anisotropy.
The location and shape of the detector represent a compromise considering the trajectories of the \btext particles with different energies and gyro-radii that depend on the strength and direction of the magnetic field they transverse.
On the one hand, \btext particles with low energy or emission angle close to the beam axis have very small gyro-radii and can pass through the aperture and thus be missed.
\btext particles with high energy or emission angle close to 90\degree to the magnetic field axis, on the other hand, show large gyro-radii, possibly larger than the size of the detector, which means they would remain undetected, too.
As a result of the simulations, the front detector is designed like a funnel that is placed in a large detector chamber in front of the magnet, (7) in Figure \ref{fig:beamline}, at a distance between $780$~mm and $580$~mm to the sample.
In this range, the field of the magnet is only~5 to 200~mT and the gyro-radii of the MeV \btext particles are in the order of several~cm. 

\begin{figure}
    \centering
    \includegraphics[width=0.8\columnwidth]{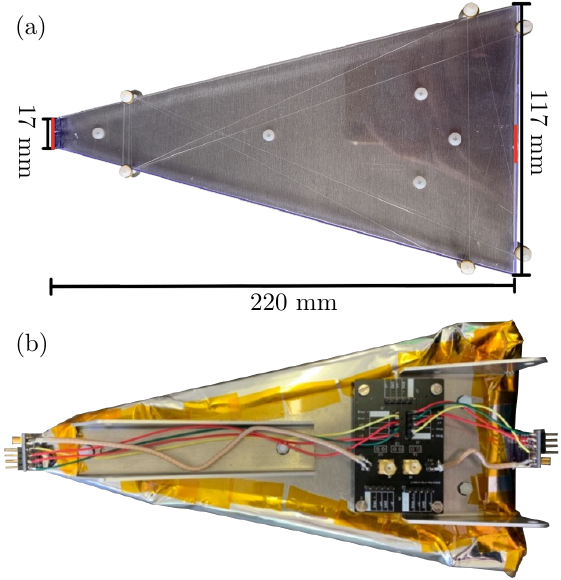}
    \caption{One of six identical elements making up the front detector. (a) Scintillator mounted on an aluminum plate during assembly. The SiPMs locations are highlighted in red. (b) Finished detector element wrapped in aluminised Mylar. The breakout PCB used for easier cable management is seen in the middle, while the PCBs with the SiPMs and amplifiers are to the left and right.}
    \label{fig:front-detector-segment}
\end{figure}

\begin{figure*}
    \centering
    \includegraphics[width=0.95\textwidth]{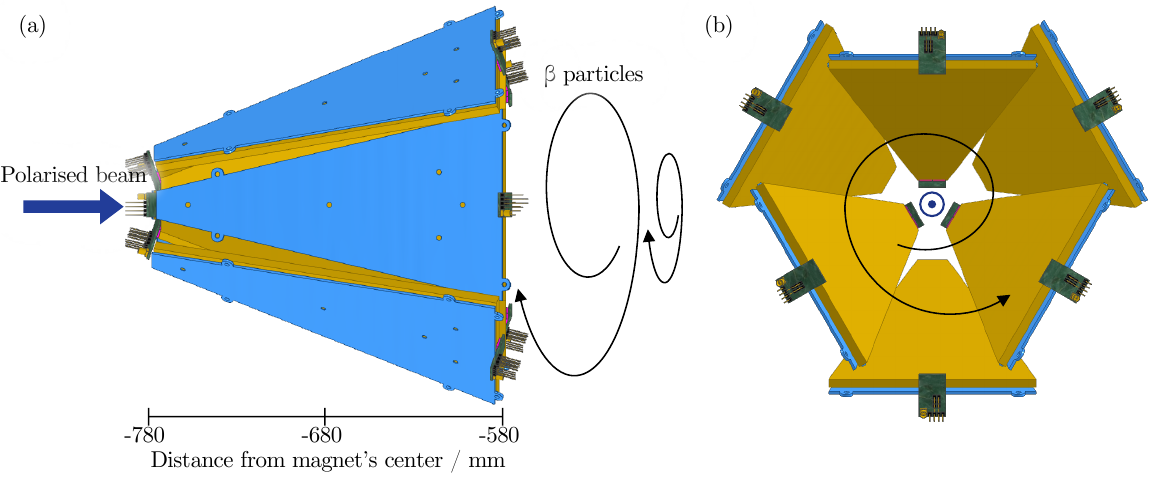}
    \caption{Schematic drawing of the front detector. (a) side view: Coming from the left, the polarised beam passes through the aperture of the detector towards the sample. The \btext particles emitted at the sample position follow a spiralling trajectory towards the detector. (b) axial view: The polarised beam comes out of the plane, passing through the aperture in the centre of the detector. The \btext particles are illustrated spiralling towards the scintillators.}
    \label{fig:front-detector-schematic}
\end{figure*}

\begin{figure}
    \centering
    \includegraphics[width=0.8\columnwidth]{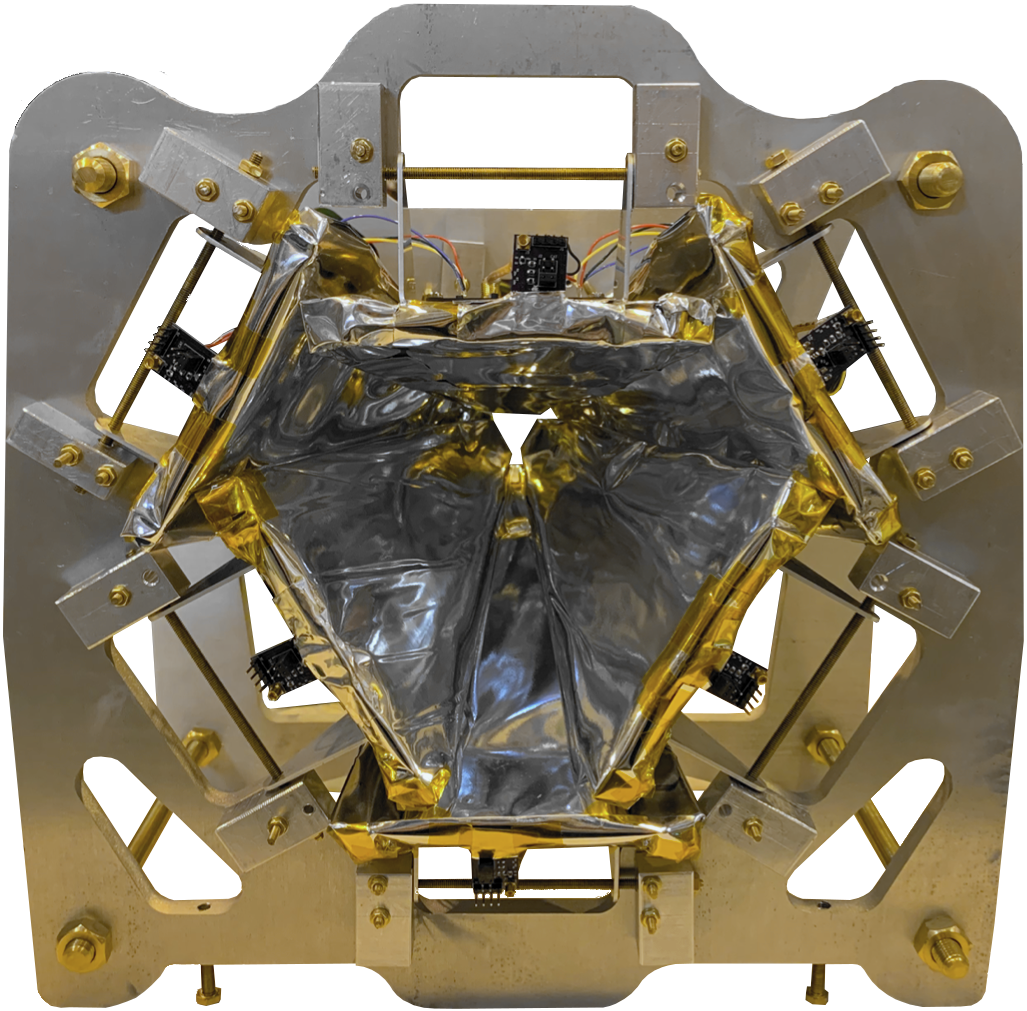}
    \caption{Front detector seen from the side of the magnet. The aperture is visible in the centre among the wrapping in aluminised Mylar and the support structure.}
    \label{fig:Front_detector}
\end{figure}

The detector consists of six trapezoidal scintillators, each with a length of 220~mm, a width between 17~mm and 117~mm and a thickness of 5~mm.
They are wrapped in aluminised Mylar and supplied with two SiPMs on opposite parallel faces, as seen in Figure \ref{fig:front-detector-segment}. 
Figure \ref{fig:front-detector-schematic} shows the design of the detector with the six scintillators being oriented as a funnel with the large opening facing the magnet.
On the other side is the hexagonal aperture with a width of 30~mm for the polarised beam to pass.
The signals from two SiPMs of one scintillator are passively summed and then actively added to the signal of the scintillator on the opposite side. 
The resulting signals of the three scintillator pairs are finally fed into the data acquisition system as three separate input channels.
Figure \ref{fig:Front_detector} depicts the final detector with its surrounding support structure.
The GEANT4 simulations of this detector geometry and position, using the magnetic field map provided by the manufacturer, show that only 20~\% of all \btext particles emitted isotropically by an unpolarised \textsuperscript{26}Na beam at the sample position are not detected.

\section{Data acquisition and control system}

The equipment upgrades and the planned studies triggered the implementation of a new control and data acquisition system (DAQ).
It allows recording spectra with a wide range of spin relaxation times from ms to~s.
To provide flexibility during the analysis, the properties of all individual \btext events in the detectors are saved in a time-resolved manner.
Moreover, the DAQ synchronises digital signals, such as the proton trigger from ISOLDE and opening the beamgate, and it controls the Doppler tuning voltage and rf excitations.
The implemented approach is twofold: a field-programmable gate array (FPGA) handles signals with critical timing, such as trigger signals, and it acquires and pre-processes the signals from the \btext detectors.
A host computer then manages the pre-processed data and peripheral devices.
Accordingly, this section describes first the requirements and the main hardware of the new system, followed by the FPGA firmware and the sequencer programme running on the host computer.

\subsection{Control and data acquisition hardware}

A variety of input and output signals are processed at the \bNMR experimental setup.
On the input side, foremost, are the analogue signals from the \btext detectors, described in Section \ref{section:Detectors}, with amplitudes of 0~V to 2~V and durations of 100~ns to 150~ns at rates of up to 1~MHz.
Analogue outputs include a -10~V to +10~V signal that is then amplified 100-fold with a Kepco BOP100 amplifier to provide the Doppler-tuning voltage of -1000~V to +1000~V.
Additional digital inputs and outputs with ns~time resolution are used to control the ISOLDE infrastructure with TTL signals.
Here, the input is a trigger of the proton beam hitting the ISOLDE target.
The outputs include the beamgate to control when the radioactive ion beam enters the \bNMR beamline, the ISOLDE cooler-buncher ISCOOL, and a laser shutter.
In addition, the DAQ communicates with several devices via USB and GPIB to read out vacuum gauges or set the rf generator parameters for NMR measurements.

To address these requirements, a combination of hardware with tailor-made software has been implemented, called VITO Control System (VCS).
The software is written in LabVIEW, building on the Control Systems++ Framework (CS++) which was developed at GSI \cite{Brand2014, Brand2016, State2022}. It consists of a collection of libraries which extends the LabVIEW-based Actor Framework.
The system uses Shared Variables together with the Data Logging and Supervisory Control Module to enable distributed and event-driven communication.
As such, it is easy to maintain and its modularity allows for hardware changes and additions.
It is published under the European Union Public Licence (EUPL).

\begin{figure}
    \centering
    \includegraphics[width=\columnwidth]{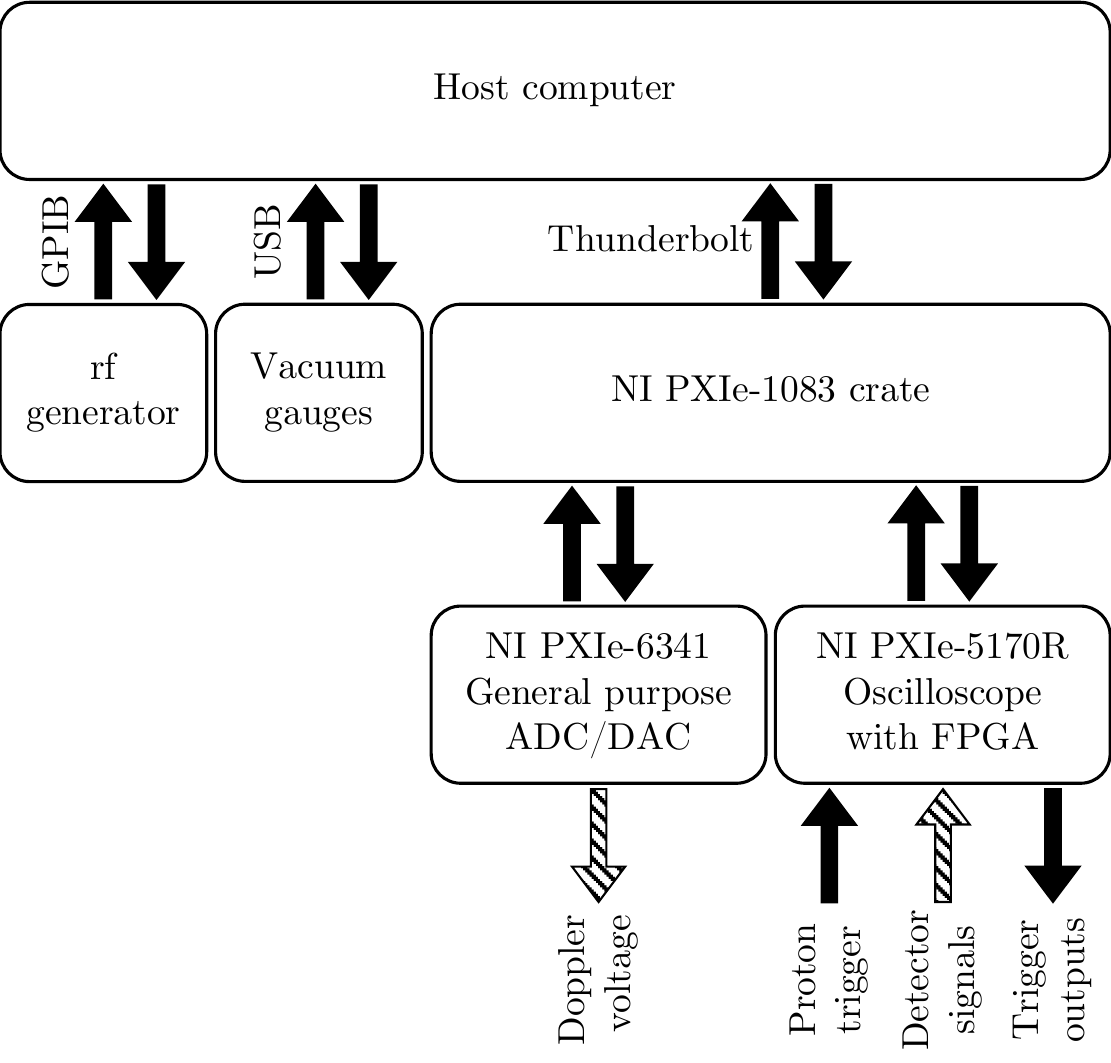}
    \caption{Schematic of the different devices, inputs and outputs handled through VCS. The analogue signals are shown with hatched arrows, all other signals are digital.}
    \label{fig:control-daq}
\end{figure}

The main hardware component of the new data acquisition is a National Instruments (NI) PXIe-5170R card. 
This 14-bit oscilloscope module offers four analogue input channels, a clock frequency of 250~MHz, a sampling rate of up to 250~MS/s and a Xilinx Kintex-7 XC7K325T FPGA that pre-processes the incoming data.
It furthermore provides the digital control signals with high temporal resolution used for the TTL signals.
The analogue and digital connections for less time-critical signals, e.g.\ the Doppler-tuning voltage, are realised with a PXIe-6341 general purpose input/output card that offers 16-bit ADCs and DACs and digital I/O at a sampling rate of 500~kS/s.
These two NI modules are installed in an NI PXIe-1083 crate that is connected to a host computer via Thunderbolt~3.
The computer itself has a GPIB extension card to communicate with the rf generator.
An overview of the different components integrated in VCS is shown schematically in Figure \ref{fig:control-daq}. 

\subsection{FPGA firmware}
\label{section:FPGA}

The firmware running on the FPGA is a modified example oscilloscope code provided by National Instruments, which has been complemented with custom features.

Its main loop runs at 125~MHz, yielding a time resolution of 8~ns.
Five digital inputs and outputs with equally high time resolution are available to generate TTL signals used for synchronising VCS with the ISOLDE infrastructure.
One of these connections receives a trigger signal once the proton pulse impinges on the ISOLDE target.
This starts the data acquisition and initialises the firmware to run its integrated pulse generator, which sends trigger signals to open the ISOLDE beamgate and the cooler buncher ISCOOL as well as a laser shutter for user-defined time windows.
All components with critical timing can thus be sequenced and controlled directly through the FPGA.

To acquire the analogue signals from the \btext detectors, options to trigger on leading and trailing edge thresholds are added.
During the measurement, the firmware on the FPGA pre-processes the detector signals to characterise each single \btext event by selected properties.
They include the time of arrival of the \btext particle with respect to the trigger, the signal amplitude, the time during which the signal is above the trigger threshold, and its corresponding integral.
Only these properties are streamed to the host computer and saved for an in-depth data analysis, rather than the raw waveforms of the signals.
This approach allows to reconstruct the signals during the analysis in post-processing while keeping the bandwidth low and thus increasing the possible signal rates, allowing for higher beam intensities.
As soon as the incoming signal amplitude exceeds the trigger threshold, the time of arrival of this \btext event is saved and its characterisation is initiated.
To determine the signal's amplitude, the incoming data samples are continuously compared to the previous ones.
For the integral, the amplitude values are summed until the signal amplitude falls below the trigger threshold.
Also, the so-called window coincidence mask is recorded, which gives information on the channels that are simultaneously active at any point during a \btext event. The aim is to identify particles that have been seen in more than one detector, which can happen due to the funnel-geometry of the front detector and the spiral trajectories of the \btext particles.
In addition, the FPGA counts the number of \btext events during specified time windows, which is then streamed to the host computer for a live view of the incoming data.
It is also possible to record the raw signal from the detectors for debugging purposes using a built-in oscilloscope function.

\subsection{Measurement sequencer}
The main application on the host computer runs a sequencer to handle the measurements and the live view of the incoming data.
A graphical interface prompts the user to choose a measurement type and to define the parameters specific to it: (1) hyperfine structure scan (HFS) with \btext asymmetry observed as a function of the Doppler-tuning voltage; (2) relaxation time measurement with \btext asymmetry observed as a function of time after proton pulse at a constant Doppler-tuning voltage; (3) \bNMR scan with \btext asymmetry observed as a function of the rf frequency at a constant Doppler-tuning voltage.
The user also defines the range and the step size for the parameter that is scanned (Doppler-tuning voltage or rf frequency), as well as the number of scans with these settings that are added to collect more statistics. 
In the case of an NMR scan, for example, it is necessary to enter the value of the Doppler-tuning voltage, as well as the the rf range and step size, modulation, and amplitude. 
Then, the program calculates the parameters of all steps covering this measurement range and stores them in a queue.
The measurement is such that the properties of all \btext events for one step are recorded for a time that is determined by the user based on the half-life of the isotope and its relaxation time.
Using the number of \btext events during user-defined time windows, the \btext decay asymmetry is calculated as the normalised difference in counts seen in the front and the back detector.
It is shown in a live view to evaluate incoming \bNMR resonances online.

\begin{figure}
    \centering
    \includegraphics[width=\columnwidth]{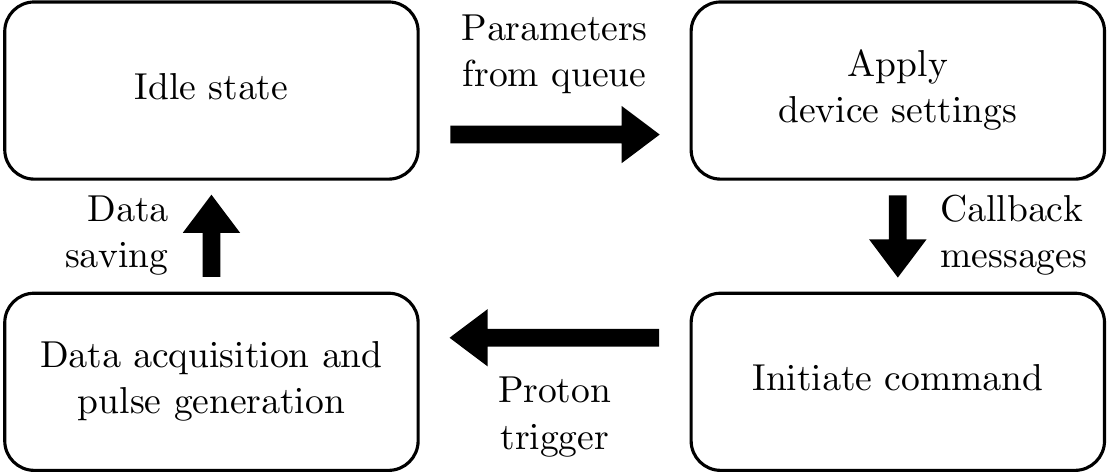}
    \caption{Schematic of the sequence to record the \btext decay asymmetry for one parameter set, starting with the FPGA being in the idle state.}
    \label{fig:FlowChart}
\end{figure}

To run the data acquisition, the sequencer processes the parameter sets from the queue to coordinate the external devices and the FPGA.
A schematic is shown in Figure \ref{fig:FlowChart}, depicting the procedure to record a single step of a scan, equivalent to one bunch of isotopes from ISOLDE.
Initially, the FPGA is in the idle state while the host applies the next parameter set.
Once the callback messages from all devices confirming the settings of the Doppler-tuning voltage, rf, etc.\ have been acknowledged, a command is sent to the FPGA to initiate it.
As soon as the FPGA receives a trigger from ISOLDE, it starts the data acquisition as well as the pulse generator, as explained in Section~\ref{section:FPGA}.
The FPGA pre-processes the incoming \btext signals while the host application immediately fetches this data from the FPGA buffer memory, so that the latter is available for the new incoming data. 
As soon as the data acquisition for this step (i.e.\ parameter set) is finished, this data is written to disk in the TDMS file format and the FPGA returns to the idle state. 
This process is repeated until there are no more parameter sets remaining in the queue, i.e.\ the last step in the last scan has been recorded).
When a measurement is completed, commands to enter a `safe mode' are sent to all connected devices, which turns all hardware outputs off until the next measurement is started.

The above features of VCS make it more suitable for the upcoming \bNMR studies compared to the previous system.
Because each single \btext event is characterised, VCS allows for more flexibility and it opens new possibilities during the data analysis, especially with respect to time-dependent behaviour of the signals, as described later in Section \ref{section:2dfits}.

\section{Beamline commissioning}

To commission the upgraded beamline, hyperfine structure scans and \bNMR spectra were acquired for \textsuperscript{26}Na ($t_{1/2}=1.1$~s, $I=3$).
To produce the beam of this short-lived isotope, a UC$_x$ target was bombarded with $3\cdot 10^{13}$ protons at an energy of 1.4~GeV \cite{Gottberg2016, Borge2016} every 3.6~s to produce about $10^7$~nuclei/s.
The atoms were surface-ionised and accelerated to an energy of 50~keV before the beam was mass-separated in the ISOLDE High Resolution Separator and it entered the \bNMR beamline.
Because Na isotopes are more easily polarised as atoms rather than ions, the ion beam was neutralised in a charge exchange cell by passing through Na vapour \cite{Gins2019}.
50~mW of circularly polarised laser light at 589~nm was used for optically pumping the $3s^2S_{1/2} \rightarrow 3p^2P_{3/2}$ transition ($D_2$ line).

The spin-polarised atom beam was then implanted into different samples: a cubic crystal or a vacuum-compatible liquid on a mica sheet.
Both types of samples are oriented to the beam in the same way as before the upgrades, see e.g.\ \cite{Croese2021}. 
The crystal used is NaF in the dimensions 10~mm by 10~mm by 1~mm, glued vertically to a 3d-printed piece that fits into the sample holder on the breadboard, see Section \ref{section:SCmagnetBreadboard}.
As a liquid sample, 10 to 20 \textmu L of EMIM-DCA (1-Ethyl-3-methyl-imidazolium dicyanamide) are used \cite{Dziubinska2021}.
Like other ionic liquids, EMIM-DCA has a very low vapour pressure that makes it suitable for vacuum conditions \cite{Bier2010}.
It is distributed over the surface of an 8-mm-diameter mica sheet oriented at 45\degree\ to the vertical on the beam axis.

The ions are let into the beamline for the first 500~ms after the proton impact on the target, while their \btext particles are recorded for 2~s. 
Since the DAQ acquires properties of all individual \btext particles, including their time of arrival, as explained in Section \ref{section:FPGA}, there are several options how to process the data and infer the \btext asymmetry. 

\subsection{Hyperfine structure scans}
\begin{figure}
    \centering
    \includegraphics[width=\columnwidth]{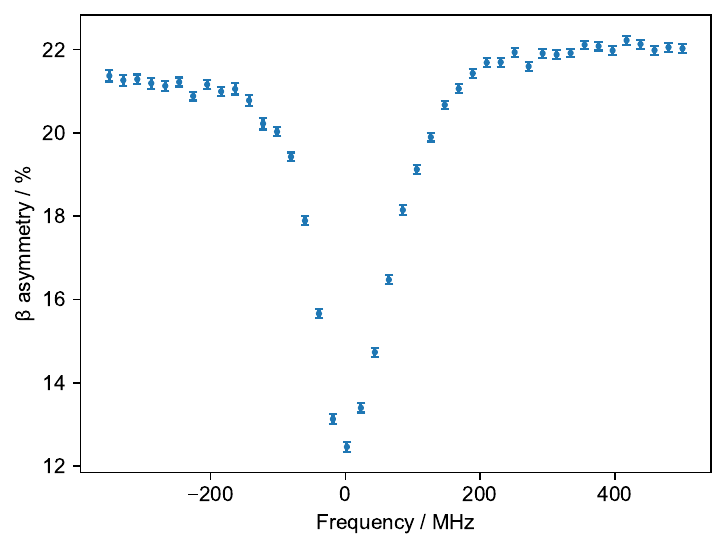}
    \caption{Part of the hyperfine structure ($D_2$ line) of \textsuperscript{26}Na  implanted into a NaF crystal, seen in \textbeta-decay asymmetry.}
    \label{fig:HFS_Na}
\end{figure}

To polarise the nuclear spin via optical pumping, the beam needs to be tuned in resonance with the laser beam, which is done by changing the Doppler-tuning voltage.
A resonance is then observed as an increase or decrease with respect to the baseline \btext decay asymmetry.
This baseline is not necessarily at 0~\% due to the instrumental asymmetry of the detector configuration. 
A part of the hyperfine structure of \textsuperscript{26}Na is shown in Figure~\ref{fig:HFS_Na}.
It indicates the highest \btext decay asymmetry that can be achieved in the $D_2$ transition, as described in detail in \cite{Kowalska2017}. 
The resonance was recorded for an acquisition duration of 2~s after polarising \textsuperscript{26}Na with \sigmap-polarised laser light and implanting it into a NaF crystal.
It shows an amplitude of 10~\% asymmetry and a high signal-to-noise ratio even though only a single scan was acquired.
Such a HFS scan was then repeated for the ionic-liquid host EMIM-DCA.
While the amplitude was smaller because of a faster relaxation in the liquid compared to the crystal, one scan suffices to obtain a resonance.

\subsection{Narrower \bNMR resonances}
\label{section:1dfits}

\begin{figure}
    \centering
    \includegraphics[width=\columnwidth]{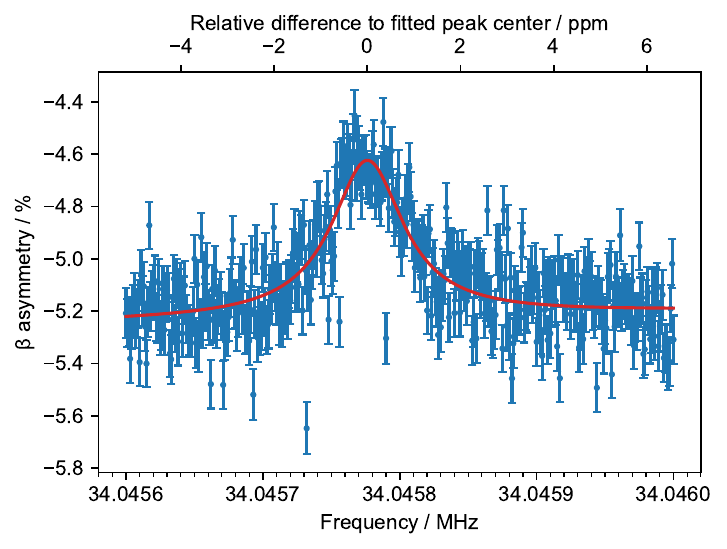}
    \caption{Example of a \bNMR spectrum of \textsuperscript{26}Na in EMIM-DCA.}
    \label{fig:NMR_Na}
\end{figure}

Having established optimum polarisation conditions with the HFS scan, several \bNMR spectra of \textsuperscript{26}Na in EMIM-DCA were recorded to evaluate the upgraded setup, one of these spectra is shown in Figure \ref{fig:NMR_Na}.
Here, the \bNMR asymmetry per rf step is calculated as the integral over all \btext events recorded during the full 2~s acquisition duration per step.
Due to a highly improved homogeneity of the magnetic field, the resonance FWHM is 64~Hz (1.9~ppm) and the uncertainty in the resonance position is 1.4~Hz ($<0.1$~ppm).
This compares to, respectively, around 200~Hz (20~ppm) and 13~Hz (1.5~ppm) using the 1.2-T electromagnet \cite{Harding2020MagneticBiology}. 
In both cases, the width is dominated by a slight rf power broadening, which was verified by recording NMR spectra with different rf power. 
The absolute precision of the measurements with the upgraded beamline is increased by a factor of~3.7 due to the higher homogeneity of the magnetic field. 
These narrower resonances, in combination with 3.9 times stronger magnetic field, provide the relative precision and thus spectral resolution that is improved by a factor of~15.
This sub-ppm precision is crucial to resolve neighbouring peaks in liquid \bNMR spectra, which is relevant in studies of the interaction of metal-ion with biomolecules.


\subsection{Two-dimensional NMR fits}
\label{section:2dfits}

The most basic way of visualising \bNMR resonances is to show the \btext decay asymmetry corresponding to all \btext events recorded in a chosen observation time as a function of the applied rf frequency, as has been done in Section~\ref{section:1dfits} in Figure~\ref{fig:NMR_Na}. 
In that approach, one frequency step of an NMR scan results in one value of \btext decay asymmetry that together with values for other steps form the basis of a one-dimensional fit of the resonance along the frequency axis.
This simple method is not optimal, as it does not take the relaxation of the signal into account.
For short relaxation times, the \btext asymmetry should be based on a shorter time window to disregard \btext events recorded after the nuclear spin polarisation has relaxed.
Choosing the observation window is always a trade-off:
On the one hand, longer observation windows contain more \btext events and in return yield smaller uncertainties of individual points. 
On the other hand, they can result in a lower amplitude of the resonance of interest if the integration time is longer than the relaxation time.

\begin{figure}
    \centering
    \includegraphics[width=\columnwidth]{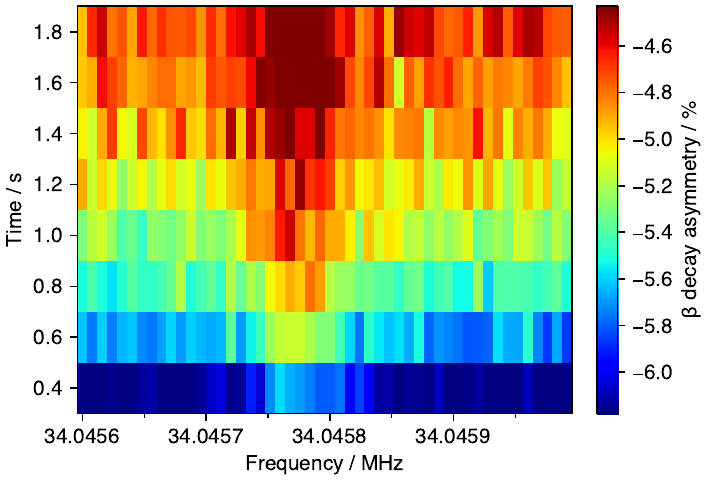}
    \caption{Two-dimensional equivalent of the previous one-dimensional \bNMR spectrum of \textsuperscript{26}Na in EMIM-DCA. It shows the \btext asymmetry as a function of the applied rf frequency and time since the implantation of the first atoms into the sample. 
    For easier visualisation, the data is binned in time steps of 0.2~s and frequency bins of 8~Hz, compared to original rf step size of 1 Hz.}
    \label{fig:2D_Na_data}
\end{figure}

Before the DAQ upgrades, a time-resolved acquisition was not possible, as the integration time had to be set in hardware. 
However, the new system, VCS, works on an event-by-event basis and records, among other properties, the time of arrival of each \btext event.
Thus, it allows processing \bNMR spectra in two dimensions, along the frequency and the time axes. 
To illustrate this more advanced approach, the \btext events of \textsuperscript{26}Na in EMIM-DCA that led to the spectrum in Figure~\ref{fig:NMR_Na} are split in 0.2-s bins to create a two-dimensional plot shown in Figure \ref{fig:2D_Na_data}. 
Such a representation can be fitted in two dimensions, thus making use of the entire data set, because it avoids the need to manually choose a time window used for one-dimensional fitting and possibly compromise the signal-to-noise ratio as would be done with one-dimensional fits.

There are three components that need to be considered when fitting the two-dimensional spectra: 
(i) The decay of polarisation and \btext decay asymmetry due to the longitudinal relaxation of the spins that reflects their interaction with the immediate environment (spin-lattice relaxation).
It is described with an exponential function with the relaxation time $T_1$ that does not depend on the applied rf frequency.
(ii) Destruction of polarisation and \btext asymmetry due to the interaction of the spins with the applied rf.
This contribution is characterised by a second time constant $T_\mathrm{L}$ which describes how much faster the polarisation is destroyed when the frequency of the applied rf signal corresponds to the Larmor frequency of the nuclei.
The frequency dependency of this contribution is represented by a Lorentzian function that describes the line shape of the resonance.
(iii) The last contribution is a constant offset that represents the instrumental \textbeta-decay asymmetry of the detectors.
The resulting fit function used to fit the two-dimensional \bNMR spectra is 

\begin{equation}
	A(\nu,t)=a e^{-t\left(\frac{1}{T_1} + \frac{w^2}{(\nu-\nu_\mathrm{L})^2+w^2} \frac{1}{T_\mathrm{L}}\right)} + b,
	\label{eq:2dfit}
\end{equation}

where $t$ is the time from the beginning of the observation window to the middle of the time bin, $a$ is the initial asymmetry generated by laser polarisation, $T_1$ is the spin-lattice relaxation time constant, $w$ is the width of the Lorentzian function, $T_\mathrm{L}$ is the additional time constant for the destruction of polarisation in resonance, $\nu$ is the frequency of the applied rf field, $\nu_\mathrm{L}$ is the Larmor frequency and $b$ is the instrumental asymmetry (i.e. \textbeta\ asymmetry observed in absence of polarisation).
The relaxation on resonance can then be described by the time constant $T_\mathrm{on}=(1/T_1+1/T_\mathrm{L})^{-1}$.

\begin{figure}
    \centering
    \includegraphics[width=\columnwidth]{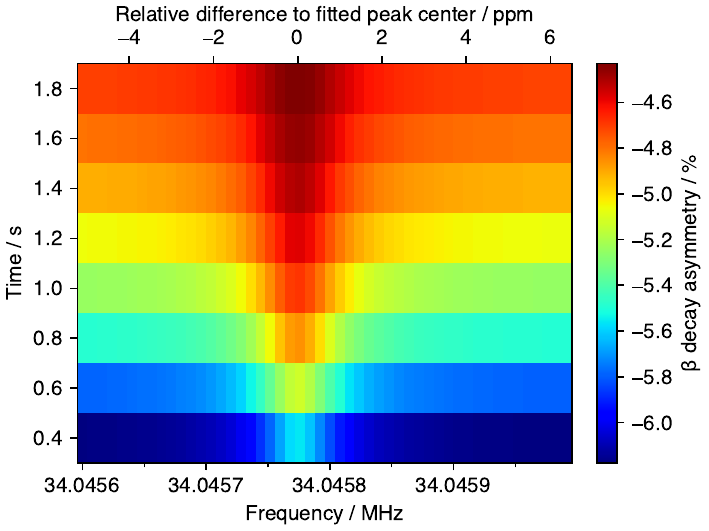}
    \caption{Two-dimensional fit of the \bNMR spectrum shown in Figure \ref{fig:2D_Na_data}.}
    \label{fig:2D_Na_fit}
\end{figure}

Only the time after closing the beamgate, i.e.\ stopping the implantation, is considered for displaying and fitting the data. 
In the data shown in Figure \ref{fig:2D_Na_data}, this time is 0.3~s.
The resulting fit, shown in Figure \ref{fig:2D_Na_fit}, yields an uncertainty in the resonance position of 2~Hz (0.06~ppm).
This represents an almost two-fold increase in the precision of the fitted resonance frequency compared to the previous one-dimensional fit.
Of equal importance as the improved precision is the elimination of a manual selection of an observation time window.
This enables a more efficient data analysis and a more complete representation of the data including relaxation.

\begin{figure}
    \centering
    \includegraphics[width=\columnwidth]{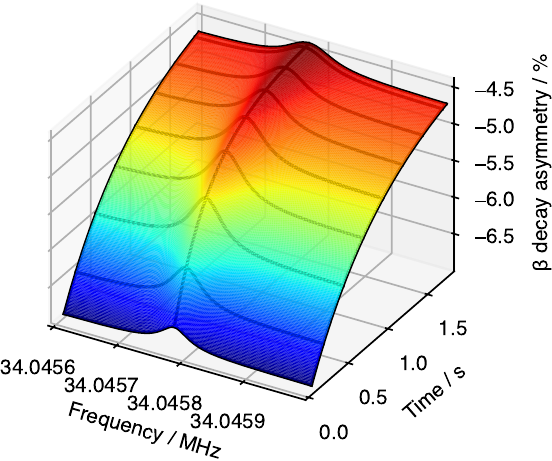}
    \caption{Three-dimensional representation of the fit of the \bNMR spectrum shown in Figure \ref{fig:2D_Na_data} illustrating the evolution of the line shape over time as the spins relax.}
    \label{fig:3D_Na_fit}
\end{figure}

In order to visualise the line shape of the \bNMR resonance as a function of time, the fit from Figure \ref{fig:2D_Na_fit} can also be represented in three dimensions, see Figure \ref{fig:3D_Na_fit}.
A projection along the time axis at the Larmor frequency shows the relaxation in the resonance condition being significantly faster than the $T_1$ relaxation off resonance.
The projections along the frequency axis for given times indicate how the resonance broadens over time as it relaxes.
This effect is otherwise not visible with the one-dimensional fits, which again highlights the advantage of the two-dimensional fits.


\begin{figure}
    \centering
    \includegraphics[width=\columnwidth]{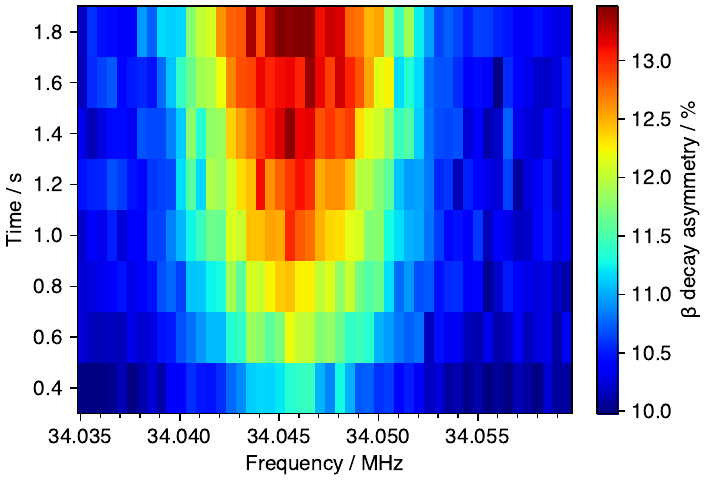}
    \caption{A two-dimensional representation of a \bNMR resonance of \textsuperscript{26}Na in a NaF crystal. The data is is binned in time steps of 0.2~s and frequency steps of 500~Hz, compared to the original 100~Hz.}
    \label{fig:2D_Na_dataNaF}
\end{figure}

\begin{figure}
    \centering
    \includegraphics[width=\columnwidth]{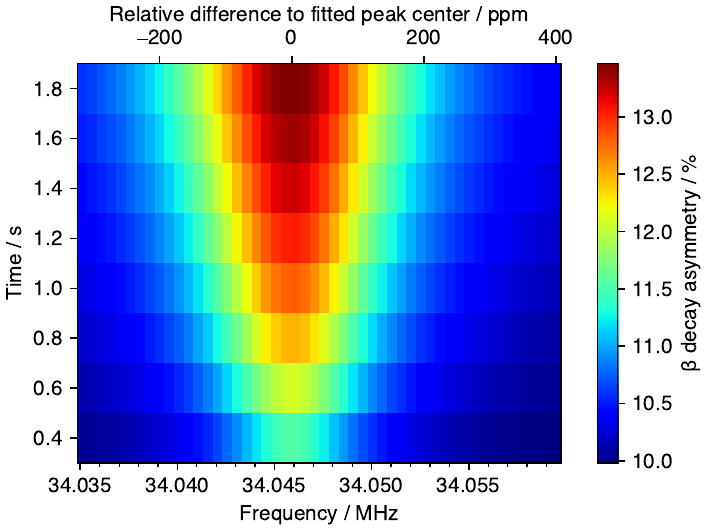}
    \caption{Two-dimensional fit of the \bNMR spectrum of \textsuperscript{26}Na in NaF shown in Figure \ref{fig:2D_Na_dataNaF}.}
    \label{fig:2D_Na_fitNaF}
\end{figure}

\begin{figure}
    \centering
    \includegraphics[width=\columnwidth]{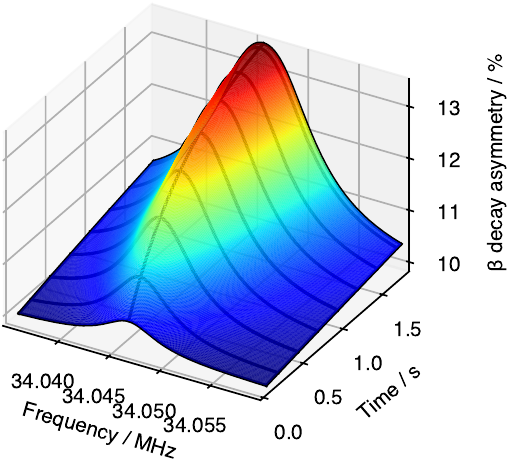}
    \caption{Three-dimensional representation of the fit of the \bNMR spectrum shown in Figure \ref{fig:2D_Na_dataNaF} illustrating the evolution of the line shape over time as the spins relax. Visible is a significantly longer $T_1$ relaxation compared to the resonance in the liquid EMIM-DCA shown in Figure~\ref{fig:3D_Na_fit}.}
    \label{fig:3D_Na_fitNaF}
\end{figure}

Analogous to the \bNMR spectrum of \textsuperscript{26}Na in the ionic liquid EMIM-DCA, Figure \ref{fig:2D_Na_dataNaF} shows a resonance of \textsuperscript{26}Na in a NaF crystal.
The corresponding fit and its three-dimensional visualisation are presented in Figure \ref{fig:2D_Na_fitNaF} and \ref{fig:3D_Na_fitNaF}.
A clear difference compared to the measurements in EMIM-DCA is the significantly longer time constant $T_1$, seen as a slower relaxation of the background in the crystal which is especially evident in the three-dimensional representation.

\subsection{Relaxation curves from NMR scans}
\begin{figure}
    \centering
    \includegraphics[width=\columnwidth]{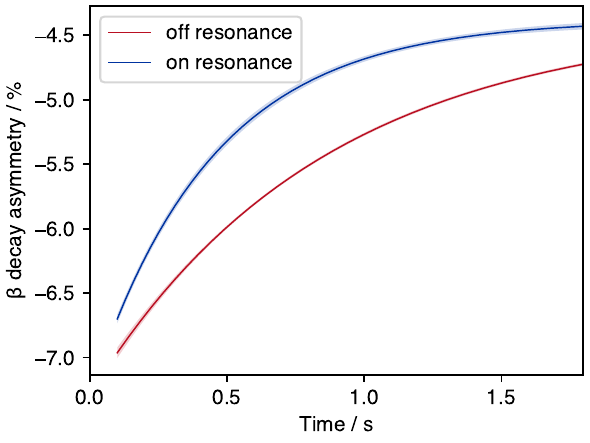}
    \caption{Relaxation curves off resonance (red), equivalent to the $T_1$ relaxation, and on resonance (blue) with the additional contribution of $T_\mathrm{L}$ extracted from a \bNMR measurement of \textsuperscript{26}Na in EMIM-DCA. The shaded areas illustrate the 1\textsigma\ uncertainties.}
    \label{fig:relaxation}
\end{figure}

Unlike conventional NMR, in which the determination of the longitudinal (spin-lattice) relaxation time $T_1$ requires multiple rf pulses, in \bNMR $T_1$ can be simply observed as the decrease of \btext decay asymmetry over time. 
During the commissioning beamtime, a dedicated measurement routine to acquire the $T_1$ relaxation time was not yet available.
However, the new data acquisition system VCS records the time of arrival for every \btext event and thus enables the two-dimensional fits along the frequency and the time axes, which were introduced in the previous section.
The $T_1$ relaxation time constant is included in the fitting function and is thus determined for every data set.
Evaluating a fit with Eq.\ \ref{eq:2dfit} for a given frequency allows to obtain a projection along the time axis.
In the case of being off resonance with the applied frequency being far from the Larmor frequency, this projection is equivalent to a $T_1$ relaxation curve.
With the applied frequency being in the range of the Larmor frequency ($\nu\approx\nu_\mathrm{L}$), the additional contribution of $T_\mathrm{L}$ is included in the projection to yield the relaxation on resonance with the time constant $T_\mathrm{on}$.
This behaviour is already visible in the three-dimensional representations of the fit in Figures \ref{fig:3D_Na_fit} and \ref{fig:3D_Na_fitNaF}.
The resulting relaxation curves on and off resonance for \textsuperscript{26}Na in EMIM-DCA are shown in Figure \ref{fig:relaxation}.
This data set is based on the \bNMR measurement shown before in Figure~\ref{fig:relaxation}, with the shaded areas indicating the 1\textsigma\ uncertainties.
The two-dimensional fits of the \bNMR resonance thus yield the relaxation curves as a by-product for every acquired spectrum.
It is therefore not necessary to acquire a dedicated relaxation measurement.

\section{Conclusions and outlook}

Described here were the upgrades performed at the \bNMR setup at the ISOLDE facility at CERN.
The principal improvement over the previous setup is a 4.7-T superconducting magnet, representing a fourfold increase in the magnetic field strength.
In addition, the magnet's sub-ppm field homogeneity over the entire measurement volume and sub-ppm temporal stability over a scan duration ensures reproducible resonances with widths in the order of a few ppm, leading to a precision in the Larmor-frequency determination of better than 1~ppm.
Furthermore, the high magnetic field and solenoidal geometry along the beam axis led to the implementation of new \btext detectors consisting of thin plastic scintillators coupled to silicon photomultipliers and arranged in an optimised geometry.
Finally, a new data acquisition system has allowed a higher degree of flexibility compared to the previous system.
It uses an FPGA to characterise every \btext event and pre-processes the data to allow for an in-depth data analysis.
The NMR spectra can now be fitted two-dimensionally along the frequency and the time axes, thus taking into account the relaxation of the polarisation for the first time.
This leads to a more representative line shape and a more precise determination of the Larmor frequency.

The above upgrades have enabled liquid \bNMR studies on the short-lived \textsuperscript{26}Na beam that reach sub-ppm precision and ns time resolution.
In the future, more devices will be added to the new DAQ VCS to supervise the experiment more comprehensively and simplify maintenance.
It is planned to move away from the TDMS file format towards HDF5, making the recorded data easier to manage and access.

Most importantly, the enhanced resolution of the setup and the upgraded data acquisition pave the way towards new applications of the \bNMR technique.
Magnetic moments can be determined with higher precision, which gives access to probing the distributions of magnetisation and neutrons inside short-lived nuclei \cite{Bissell2023}.
At the same time, the accessible small differences in the Larmor frequencies of short-lived metal ions due to their environment will allow for biochemistry studies of metal ion interaction with DNA or proteins \cite{Karg2020, Karg2022}.

\section{Acknowledgements}
This work was supported by the European Research Council (Starting Grant 640465),
and the Wolfgang Gentner Programme of the German Federal Ministry of Education and Research (13E18CHA).
We would like to thank the ISOLDE technical team for their support in installing the superconducting magnet in the ISOLDE hall and during beamtimes, the ISOLDE Collaboration for their support and the COLLAPS collaboration for the use of their tunable laser and optics.

\bibliographystyle{Template/model1-num-names}
\bibliography{References}

\begin{thebibliography}{26}
\expandafter\ifx\csname natexlab\endcsname\relax\def\natexlab#1{#1}\fi
\providecommand{\url}[1]{\texttt{#1}}
\providecommand{\href}[2]{#2}
\providecommand{\path}[1]{#1}
\providecommand{\DOIprefix}{doi:}
\providecommand{\ArXivprefix}{arXiv:}
\providecommand{\URLprefix}{URL: }
\providecommand{\Pubmedprefix}{pmid:}
\providecommand{\doi}[1]{\href{http://dx.doi.org/#1}{\path{#1}}}
\providecommand{\Pubmed}[1]{\href{pmid:#1}{\path{#1}}}
\providecommand{\bibinfo}[2]{#2}
\ifx\xfnm\relax \def\xfnm[#1]{\unskip,\space#1}\fi
\bibitem[{Mazzola et~al.(2003)Mazzola, Lambert, and Holland}]{Mazzola03}
\bibinfo{author}{E.~P. Mazzola}, \bibinfo{author}{J.~B. Lambert},
  \bibinfo{author}{L.~N. Holland}, \bibinfo{title}{Nuclear Magnetic Resonance
  Spectroscopy: An Introduction to Principles, Applications, and Experimental
  Methods}, \bibinfo{publisher}{Clarendon Press}, \bibinfo{year}{2003}.
\bibitem[{Abragam(1989)}]{Abragam1989}
\bibinfo{author}{A.~Abragam}, \bibinfo{title}{The principles of nuclear
  magnetism}, \bibinfo{publisher}{Clarendon Press}, \bibinfo{year}{1989}.
\bibitem[{Levitt(2001)}]{Levitt:500323}
\bibinfo{author}{M.~H. Levitt}, \bibinfo{title}{{Spin dynamics: basics of
  nuclear magnetic resonance}}, \bibinfo{publisher}{Wiley},
  \bibinfo{address}{New York, NY}, \bibinfo{year}{2001}. \URLprefix
  \url{https://cds.cern.ch/record/500323}.
\bibitem[{et~al.'(2023)}]{Ellis2023}
\bibinfo{author}{J.~E. et~al.'},
\newblock \bibinfo{title}{Spin hyperpolarization in modern magnetic resonance},
\newblock \bibinfo{journal}{Chemical Reviews} \bibinfo{volume}{123}
  (\bibinfo{year}{2023}) \bibinfo{pages}{1417}.
\bibitem[{Wu et~al.(1957)Wu, Ambler, Hayward, Hoppes, and Hudson}]{Wu1957}
\bibinfo{author}{C.~S. Wu}, \bibinfo{author}{E.~Ambler}, \bibinfo{author}{R.~W.
  Hayward}, \bibinfo{author}{D.~D. Hoppes}, \bibinfo{author}{R.~P. Hudson},
\newblock \bibinfo{title}{{Experimental Test of Parity Conservation in Beta
  Decay}},
\newblock \bibinfo{journal}{Phys. Rev.} \bibinfo{volume}{105}
  (\bibinfo{year}{1957}) \bibinfo{pages}{1413--1415}.
\bibitem[{Kowalska et~al.(2008)Kowalska, Yordanov, Blaum, Himpe, Lievens,
  Mallion, Neugart, Neyens, and Vermeulen}]{Kowalska2008}
\bibinfo{author}{M.~Kowalska}, \bibinfo{author}{D.~T. Yordanov},
  \bibinfo{author}{K.~Blaum}, \bibinfo{author}{P.~Himpe},
  \bibinfo{author}{P.~Lievens}, \bibinfo{author}{S.~Mallion},
  \bibinfo{author}{R.~Neugart}, \bibinfo{author}{G.~Neyens},
  \bibinfo{author}{N.~Vermeulen},
\newblock \bibinfo{title}{{Nuclear ground-state spins and magnetic moments of
  $^{27}$Mg, $^{29}$Mg, and $^{31}$Mg}},
\newblock \bibinfo{journal}{Physical Review C} \bibinfo{volume}{77}
  (\bibinfo{year}{2008}) \bibinfo{pages}{034307}.
\bibitem[{Keim et~al.(2000)Keim, Georg, Klein, Neugart, Neuroth, Wilbert,
  Lievens, Vermeeren, Brown, and Collaboration}]{Keim2000}
\bibinfo{author}{M.~Keim}, \bibinfo{author}{U.~Georg},
  \bibinfo{author}{A.~Klein}, \bibinfo{author}{R.~Neugart},
  \bibinfo{author}{M.~Neuroth}, \bibinfo{author}{S.~Wilbert},
  \bibinfo{author}{P.~Lievens}, \bibinfo{author}{L.~Vermeeren},
  \bibinfo{author}{B.~A. Brown}, \bibinfo{author}{I.~Collaboration},
\newblock \bibinfo{title}{{Measurement of the electric quadrupole moments of
  26–29Na}},
\newblock \bibinfo{journal}{The European Physical Journal A}
  \bibinfo{volume}{8} (\bibinfo{year}{2000}) \bibinfo{pages}{31--40}.
\bibitem[{Yordanov et~al.(2019)Yordanov, Kowalska, Blaum, Rydt, Flanagan,
  Himpe, Lievens, Mallion, Neugart, Neyens, Vermeulen, and
  Stroke}]{Yordanov2019}
\bibinfo{author}{D.~T. Yordanov}, \bibinfo{author}{M.~Kowalska},
  \bibinfo{author}{K.~Blaum}, \bibinfo{author}{M.~D. Rydt},
  \bibinfo{author}{K.~T. Flanagan}, \bibinfo{author}{P.~Himpe},
  \bibinfo{author}{P.~Lievens}, \bibinfo{author}{S.~Mallion},
  \bibinfo{author}{R.~Neugart}, \bibinfo{author}{G.~Neyens},
  \bibinfo{author}{N.~Vermeulen}, \bibinfo{author}{H.~Stroke},
\newblock \bibinfo{title}{{Quadrupole Moments of 29Mg and 33Mg}},
\newblock \bibinfo{journal}{Hyperfine Interactions}  (\bibinfo{year}{2019}).
\bibitem[{Sugihara et~al.(2017)Sugihara, Mihara, Shimaya, Matsuta, Fukuda,
  Ohno, Tanaka, Yamaoka, Watanabe, Iwakiri, Yanagihara, Tanaka, Du, Onishi,
  Kambayashi, Minamisono, Nishimura, Izumikawa, Ozawa, Ishibashi, Kitagawa,
  Sato, Torikoshi, and Momota}]{Sugihara2017}
\bibinfo{author}{T.~Sugihara}, \bibinfo{author}{M.~Mihara},
  \bibinfo{author}{J.~Shimaya}, \bibinfo{author}{K.~Matsuta},
  \bibinfo{author}{M.~Fukuda}, \bibinfo{author}{J.~Ohno},
  \bibinfo{author}{M.~Tanaka}, \bibinfo{author}{S.~Yamaoka},
  \bibinfo{author}{K.~Watanabe}, \bibinfo{author}{S.~Iwakiri},
  \bibinfo{author}{R.~Yanagihara}, \bibinfo{author}{Y.~Tanaka},
  \bibinfo{author}{H.~Du}, \bibinfo{author}{K.~Onishi},
  \bibinfo{author}{S.~Kambayashi}, \bibinfo{author}{T.~Minamisono},
  \bibinfo{author}{D.~Nishimura}, \bibinfo{author}{T.~Izumikawa},
  \bibinfo{author}{A.~Ozawa}, \bibinfo{author}{Y.~Ishibashi},
  \bibinfo{author}{A.~Kitagawa}, \bibinfo{author}{S.~Sato},
  \bibinfo{author}{M.~Torikoshi}, \bibinfo{author}{S.~Momota},
\newblock \bibinfo{title}{Nmr detection of short-lived β-emitter 12n implanted
  in water},
\newblock \bibinfo{journal}{Hyperfine Interactions} \bibinfo{volume}{238}
  (\bibinfo{year}{2017}).
\bibitem[{Mihara et~al.(2019)Mihara, Sugihara, Fukuda, Homma, Izumikawa,
  Kitagawa, Matsuta, Minaisono, Momota, Nagatomo, Nishibata, Nishimura,
  Ohnishi, Ohtsubo, Ozawa, Sato, Tanaka, Wakabayashi, Yagi, and
  Yanagihara}]{Mihara2019}
\bibinfo{author}{M.~Mihara}, \bibinfo{author}{T.~Sugihara},
  \bibinfo{author}{M.~Fukuda}, \bibinfo{author}{A.~Homma},
  \bibinfo{author}{T.~Izumikawa}, \bibinfo{author}{A.~Kitagawa},
  \bibinfo{author}{K.~Matsuta}, \bibinfo{author}{T.~Minaisono},
  \bibinfo{author}{S.~Momota}, \bibinfo{author}{T.~Nagatomo},
  \bibinfo{author}{H.~Nishibata}, \bibinfo{author}{D.~Nishimura},
  \bibinfo{author}{K.~Ohnishi}, \bibinfo{author}{T.~Ohtsubo},
  \bibinfo{author}{A.~Ozawa}, \bibinfo{author}{S.~Sato},
  \bibinfo{author}{M.~Tanaka}, \bibinfo{author}{R.~Wakabayashi},
  \bibinfo{author}{S.~Yagi}, \bibinfo{author}{R.~Yanagihara},
\newblock \bibinfo{title}{Beta-nmr of short-lived nucleus 17n in liquids},
\newblock \bibinfo{journal}{Hyperfine Interactions} \bibinfo{volume}{240}
  (\bibinfo{year}{2019}).
\bibitem[{Croese et~al.(2021)Croese, Baranowski, Bissell, Dziubinska-K\"uhn,
  Gins, Harding, Jolivet, Kanellakopoulos, Karg, Kulesz, Madurga~Flores,
  Neyens, Pallada, Pietrzyk, Pomorski, Wagenknecht, Zakoucky, and
  Kowalska}]{Croese2021}
\bibinfo{author}{J.~Croese}, \bibinfo{author}{M.~Baranowski},
  \bibinfo{author}{M.~L. Bissell}, \bibinfo{author}{K.~M. Dziubinska-K\"uhn},
  \bibinfo{author}{W.~Gins}, \bibinfo{author}{R.~D. Harding},
  \bibinfo{author}{R.~B. Jolivet}, \bibinfo{author}{A.~Kanellakopoulos},
  \bibinfo{author}{B.~Karg}, \bibinfo{author}{K.~Kulesz},
  \bibinfo{author}{M.~Madurga~Flores}, \bibinfo{author}{G.~Neyens},
  \bibinfo{author}{S.~Pallada}, \bibinfo{author}{R.~Pietrzyk},
  \bibinfo{author}{M.~Pomorski}, \bibinfo{author}{P.~Wagenknecht},
  \bibinfo{author}{D.~Zakoucky}, \bibinfo{author}{M.~Kowalska},
\newblock \bibinfo{title}{{High-accuracy liquid-sample \textbeta-NMR setup at
  ISOLDE}},
\newblock \bibinfo{journal}{Nuclear Instruments and Methods in Physics Research
  Section A: Accelerators, Spectrometers, Detectors and Associated Equipment}
  \bibinfo{volume}{1020} (\bibinfo{year}{2021}) \bibinfo{pages}{165862}.
\bibitem[{Bohr and Weisskopf(1950)}]{BohrWeisskopf1950}
\bibinfo{author}{A.~Bohr}, \bibinfo{author}{V.~F. Weisskopf},
\newblock \bibinfo{title}{The influence of nuclear structure on the hyperfine
  structure of heavy elements},
\newblock \bibinfo{journal}{Phys. Rev.} \bibinfo{volume}{77}
  (\bibinfo{year}{1950}) \bibinfo{pages}{94--98}.
\bibitem[{{M. Bissell, M. Kowalska, and others}(2023)}]{Bissell2023}
\bibinfo{author}{{M. Bissell, M. Kowalska, and others}},
\newblock \bibinfo{title}{Magnetic moment of 11 be with ppm accuracy},
\newblock \bibinfo{journal}{INTC proposal, CERN--INTC-2023-014, INTC-P-655}
  (\bibinfo{year}{2023}).
\bibitem[{{Jancso et al.}(2017)}]{Jancso2017}
\bibinfo{author}{{Jancso et al.}},
\newblock \bibinfo{title}{{Exploring solid state physics properties with
  radioactive isotopes TDPAC and $\beta$-NMR applications in chemistry and
  biochemistry}},
\newblock \bibinfo{journal}{Journal of Physics G} \bibinfo{volume}{44}
  (\bibinfo{year}{2017}) \bibinfo{pages}{064003}.
\bibitem[{Harding et~al.(2020)Harding, Pallada, Croese, Antu{\v{s}}ek,
  Baranowski, Bissell, Cerato, Dziubinska-K{\"{u}}hn, Gins, Gustafsson, Javaji,
  Jolivet, Kanellakopoulos, Karg, Kempka, Kocman, Kozak, Kulesz, Flores,
  Neyens, Pietrzyk, Plavec, Pomorski, Skrzypczak, Wagenknecht, Wienholtz,
  Wolak, Xu, Zakoucky, and Kowalska}]{Harding2020MagneticBiology}
\bibinfo{author}{R.~D. Harding}, \bibinfo{author}{S.~Pallada},
  \bibinfo{author}{J.~Croese}, \bibinfo{author}{A.~Antu{\v{s}}ek},
  \bibinfo{author}{M.~Baranowski}, \bibinfo{author}{M.~L. Bissell},
  \bibinfo{author}{L.~Cerato}, \bibinfo{author}{K.~M. Dziubinska-K{\"{u}}hn},
  \bibinfo{author}{W.~Gins}, \bibinfo{author}{F.~P. Gustafsson},
  \bibinfo{author}{A.~Javaji}, \bibinfo{author}{R.~B. Jolivet},
  \bibinfo{author}{A.~Kanellakopoulos}, \bibinfo{author}{B.~Karg},
  \bibinfo{author}{M.~Kempka}, \bibinfo{author}{V.~Kocman},
  \bibinfo{author}{M.~Kozak}, \bibinfo{author}{K.~Kulesz},
  \bibinfo{author}{M.~M. Flores}, \bibinfo{author}{G.~Neyens},
  \bibinfo{author}{R.~Pietrzyk}, \bibinfo{author}{J.~Plavec},
  \bibinfo{author}{M.~Pomorski}, \bibinfo{author}{A.~Skrzypczak},
  \bibinfo{author}{P.~Wagenknecht}, \bibinfo{author}{F.~Wienholtz},
  \bibinfo{author}{J.~Wolak}, \bibinfo{author}{Z.~Xu},
  \bibinfo{author}{D.~Zakoucky}, \bibinfo{author}{M.~Kowalska},
\newblock \bibinfo{title}{{Magnetic Moments of Short-Lived Nuclei with
  Part-per-Million Accuracy: Toward Novel Applications of {\ss} -Detected NMR
  in Physics, Chemistry, and Biology}},
\newblock \bibinfo{journal}{Physical Review X} \bibinfo{volume}{10}
  (\bibinfo{year}{2020}) \bibinfo{pages}{41061}.
\bibitem[{{B. Karg, M. Kowalska, and others}(2020)}]{Karg2020}
\bibinfo{author}{{B. Karg, M. Kowalska, and others}},
\newblock \bibinfo{title}{Liquid $\beta$-nmr studies of the interaction of na
  and k cations with dna g-quadruplex structures},
\newblock \bibinfo{journal}{INTC proposal, CERN-INTC-2020-034 / INTC-P-560}
  (\bibinfo{year}{2020}) \bibinfo{pages}{1}.
\bibitem[{Gins et~al.(2019)Gins, Harding, Baranowski, Bissell, Ruiz, Kowalska,
  Neyens, Pallada, Severijns, Velten, Wienholtz, Xu, Yang, and
  Zakoucky}]{Gins2019}
\bibinfo{author}{W.~Gins}, \bibinfo{author}{R.~D. Harding},
  \bibinfo{author}{M.~Baranowski}, \bibinfo{author}{M.~L. Bissell},
  \bibinfo{author}{R.~F.~G. Ruiz}, \bibinfo{author}{M.~Kowalska},
  \bibinfo{author}{G.~Neyens}, \bibinfo{author}{S.~Pallada},
  \bibinfo{author}{N.~Severijns}, \bibinfo{author}{P.~Velten},
  \bibinfo{author}{F.~Wienholtz}, \bibinfo{author}{Z.~Y. Xu},
  \bibinfo{author}{X.~F. Yang}, \bibinfo{author}{D.~Zakoucky},
\newblock \bibinfo{title}{{A new beamline for laser spin-polarization at
  ISOLDE}},
\newblock \bibinfo{journal}{Nuclear Instruments and Methods in Physics Research
  Section A: Accelerators, Spectrometers, Detectors and Associated Equipment}
  \bibinfo{volume}{925} (\bibinfo{year}{2019}) \bibinfo{pages}{24--32}.
\bibitem[{Kowalska et~al.(2017)Kowalska, Aschenbrenner, Baranowski, Bissell,
  Gins, Harding, Heylen, Neyens, Pallada, Severijns, Velten, Walczak,
  Wienholtz, Xu, Yang, and Zakoucky}]{Kowalska2017}
\bibinfo{author}{M.~Kowalska}, \bibinfo{author}{P.~Aschenbrenner},
  \bibinfo{author}{M.~Baranowski}, \bibinfo{author}{M.~L. Bissell},
  \bibinfo{author}{W.~Gins}, \bibinfo{author}{R.~D. Harding},
  \bibinfo{author}{H.~Heylen}, \bibinfo{author}{G.~Neyens},
  \bibinfo{author}{S.~Pallada}, \bibinfo{author}{N.~Severijns},
  \bibinfo{author}{P.~Velten}, \bibinfo{author}{M.~Walczak},
  \bibinfo{author}{F.~Wienholtz}, \bibinfo{author}{Z.~Y. Xu},
  \bibinfo{author}{X.~F. Yang}, \bibinfo{author}{D.~Zakoucky},
\newblock \bibinfo{title}{{New laser polarization line at the ISOLDE
  facility}},
\newblock \bibinfo{journal}{Journal of Physics G: Nuclear and Particle Physics}
  \bibinfo{volume}{44} (\bibinfo{year}{2017}) \bibinfo{pages}{084005}.
\bibitem[{Brand and Neidherr(2014)}]{Brand2014}
\bibinfo{author}{H.~Brand}, \bibinfo{author}{D.~Neidherr},
  \bibinfo{title}{Status of the cs framework and its successor cs++},
  \bibinfo{year}{2014}. \DOIprefix\doi{10.15120/GR-2015-1-FG-GENERAL-41}.
\bibitem[{Brand and Neidherr(2016)}]{Brand2016}
\bibinfo{author}{H.~Brand}, \bibinfo{author}{D.~Neidherr}, \bibinfo{title}{Cs++
  – ni actor framework-based class library}, \bibinfo{year}{2016}.
\bibitem[{State et~al.(2022)State, Beck, Amanbayev, Balabanski, Brand,
  Constantin, Dickel, Hornung, Nichita, Plaß, Roesch, Rotaru, Scheidenberger,
  Siebring, Spataru, Tortorelli, and Zhao}]{State2022}
\bibinfo{author}{A.~State}, \bibinfo{author}{S.~Beck},
  \bibinfo{author}{D.~Amanbayev}, \bibinfo{author}{D.~Balabanski},
  \bibinfo{author}{H.~Brand}, \bibinfo{author}{P.~Constantin},
  \bibinfo{author}{T.~Dickel}, \bibinfo{author}{C.~Hornung},
  \bibinfo{author}{D.~Nichita}, \bibinfo{author}{W.~Plaß},
  \bibinfo{author}{H.~Roesch}, \bibinfo{author}{A.~Rotaru},
  \bibinfo{author}{C.~Scheidenberger}, \bibinfo{author}{J.~Siebring},
  \bibinfo{author}{A.~Spataru}, \bibinfo{author}{N.~Tortorelli},
  \bibinfo{author}{J.~Zhao},
\newblock \bibinfo{title}{The slow control system of the frs ion catcher},
\newblock \bibinfo{journal}{Nuclear Instruments and Methods in Physics Research
  Section A: Accelerators, Spectrometers, Detectors and Associated Equipment}
  \bibinfo{volume}{1034} (\bibinfo{year}{2022}) \bibinfo{pages}{166772}.
\bibitem[{Gottberg(2016)}]{Gottberg2016}
\bibinfo{author}{A.~Gottberg},
\newblock \bibinfo{title}{Target materials for exotic {ISOL} beams},
\newblock \bibinfo{journal}{Nuclear Instruments and Methods in Physics Research
  Section B: Beam Interactions with Materials and Atoms} \bibinfo{volume}{376}
  (\bibinfo{year}{2016}) \bibinfo{pages}{8--15}.
\bibitem[{Borge(2016)}]{Borge2016}
\bibinfo{author}{M.~Borge},
\newblock \bibinfo{title}{Highlights of the {ISOLDE} facility and the
  {HIE}-{ISOLDE} project},
\newblock \bibinfo{journal}{Nuclear Instruments and Methods in Physics Research
  Section B: Beam Interactions with Materials and Atoms} \bibinfo{volume}{376}
  (\bibinfo{year}{2016}) \bibinfo{pages}{408--412}.
\bibitem[{Dziubinska-Kühn et~al.(2021)Dziubinska-Kühn, Croese, Pupier,
  Matysik, Viger-Gravel, Karg, and Kowalska}]{Dziubinska2021}
\bibinfo{author}{K.~Dziubinska-Kühn}, \bibinfo{author}{J.~Croese},
  \bibinfo{author}{M.~Pupier}, \bibinfo{author}{J.~Matysik},
  \bibinfo{author}{J.~Viger-Gravel}, \bibinfo{author}{B.~Karg},
  \bibinfo{author}{M.~Kowalska},
\newblock \bibinfo{title}{Structural analysis of water in ionic liquid domains
  – a low pressure study},
\newblock \bibinfo{journal}{Journal of Molecular Liquids} \bibinfo{volume}{334}
  (\bibinfo{year}{2021}) \bibinfo{pages}{116447}.
\bibitem[{Bier and Dietrich(2010)}]{Bier2010}
\bibinfo{author}{M.~Bier}, \bibinfo{author}{S.~Dietrich},
\newblock \bibinfo{title}{{Vapour pressure of ionic liquids}},
\newblock \bibinfo{journal}{Molecular Physics} \bibinfo{volume}{108}
  (\bibinfo{year}{2010}) \bibinfo{pages}{211--214}.
\bibitem[{Karg and Kowalska(2022)}]{Karg2022}
\bibinfo{author}{B.~Karg}, \bibinfo{author}{M.~Kowalska},
\newblock \bibinfo{title}{Liquid $\beta$-nmr studies of the interaction of na
  and k cations with dna g-quadruplex structures},
\newblock \bibinfo{journal}{INTC proposal, CERN--INTC-2022-001,
  INTC-P-560-ADD-1}  (\bibinfo{year}{2022}).

\end{thebibliography}

\end{document}